# Meter-scale supersonic gas jets for multi-GeV laser-plasma accelerators

B. Miao, J. E. Shrock, E. Rockafellow, A. Sloss and H. M. Milchberg

*Institute for Research in Electronics and Applied Physics, University of Maryland, College Park, Maryland 20742, USA*

**Abstract:** Pushing the high energy frontier of laser wakefield electron acceleration (LWFA) to 10 GeV and beyond requires extending the propagation of relativistic intensity pulses to ~1 m in a low density ($N_e \sim 10^{17} \text{cm}^{-3}$) plasma waveguide. We present the development and characterization of two types of supersonic gas jet for meter-scale multi-GeV laser wakefield accelerators. The first type is a 30-cm long single-module gas jet, which demonstrates good axial uniformity using hydrogen, the preferred working gas for LWFA. The second type is a modular jet composed of multiple 11-cm-long modules. Longitudinal density profile control is demonstrated with a 2-module (22 cm long) hydrogen jet using gas valve trigger timing. A 1.0-m-long jet is then assembled from 9 modules, and generation of 1.0-m long hydrogen plasma is demonstrated using a femtosecond Bessel beam. To our knowledge, this is the longest gas jet laser plasma yet generated.

## I. Introduction

Laser wakefield acceleration (LWFA) is a promising technique for compact electron accelerators due to the high accelerating gradient supported by plasma [1,2], typically $\sim 10^3$ greater than in conventional rf-driven linear accelerators. Numerous applications of LWFA have been demonstrated, including high energy photon sources [3,4], secondary particle generation [5,6] and free-electron lasers [7]. For high energy colliders, staging of multiple LWFAs is proposed [8]. Recent advances in laser technology and computer simulations have enabled the generation of multi-GeV electron beams [9–14], but it remains a challenge to demonstrate a combination of high charge (nC) and low energy spread (sub-percent). Precise control of the laser driver and plasma target are essential to achieve these goals.

In order to reach ~10 GeV energy gain in a single LWFA stage without dephasing of the acceleration process [2], a laser pulse must maintain its normalized vector potential $a_0 > 1$ by optical guiding in a low-density meter-scale plasma waveguide with on-axis density $N_e \sim 10^{17} \text{cm}^{-3}$ [10]. Development of multi-GeV LWFAs imposes demanding requirements on a plasma target: it should be easily accessible by experimental diagnostics, free standing to avoid laser and plasma damage of nearby materials, amenable to high repetition rate operation, and programmable in density profile. Well-known plasma targets in which waveguides are generated, namely gas cells [15–17] and capillary discharges [11,18,19], cannot meet all of these requirements. All requirements, however, can be met by an appropriately designed gas jet.

In this paper, we present the development of two types of meter-scale, low density, supersonic gas jet. The first type is a 30-cm-long single-module jet applied in recent research to deliver accelerated electron beams up to 10 GeV [10,20] using hydrogen as the working gas. The second type is a modular jet, with which we demonstrate hydrogen plasma waveguides up to 1.0 m long. A modular jet enables axial density tailoring to mitigate dephasing between the accelerated electron beam and the optically guided laser pulse [16,21–23]. It also enables control of electron injection along the laser propagation path, especially ionization injection [24,25] using specially selected gas species.

## II. 30-cm single-module supersonic gas jet

Supersonic gas jets are widely used in laser plasma interactions [26–33] with a typical range of gas density from $\sim 10^{16}$ cm$^{-3}$ to $\sim 10^{21}$ cm$^{-3}$. Most supersonic gas jets provide a gas plume up to a couple of centimeters long and are backed by a single gas valve. For multi-GeV LWFAs, the laser propagation distance in plasma must be at least tens of centimeters. Such plasma is generated in a gas sheet tens of centimeters long [10,20], for which it is difficult for a single inlet and valve to maintain the needed density along with a fast risetime. The latter is especially needed so that the working density is established quickly without deleterious gas loading of the experimental vacuum chamber.

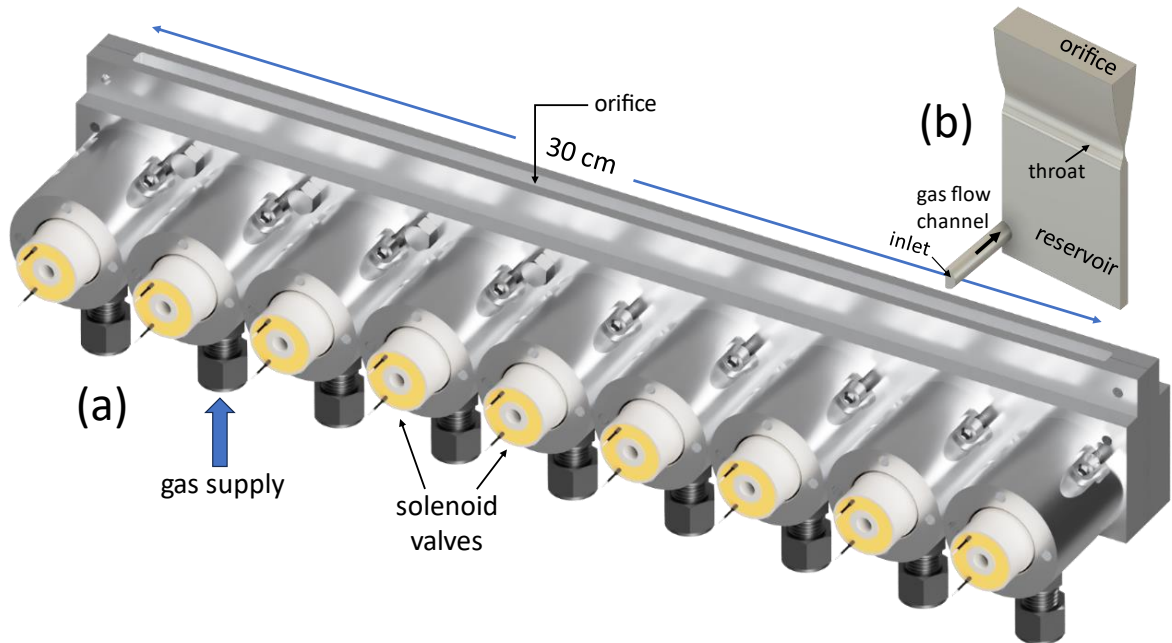

**Figure 1. (a)** 30-cm-long single-module supersonic slit nozzle, fed by 9 solenoid valves, each supplied by separate or manifold-linked gas supply lines. **(b)** Longitudinal section of internal volume of jet, showing valve-actuated gas flow to reservoir, whose role is to axially smooth the gas flow. This section also serves as the elemental volume for our fluid flow simulations. The axially-extended supersonic nozzle, with a throat width of 200 μm, is designed for Mach 5 flow.

Our single-module 30 cm-long gas jet design consists of an elongated supersonic slit nozzle fed by a reservoir backed by multiple solenoid valves. A prior 20 cm long version was used in LWFA experiments achieving electron bunches up to 5 GeV [10,14,20]. A drawing of the jet is shown in Fig. 1. The reservoir, which serves to smooth the effect of discrete valves on axial variation of nozzle flow, is fed by 9 side-mounted miniature solenoid valves (Parker C21), each in a specialized coupler. The jet is composed of two blocks, with the valves mounted on the front block most visible in the figure. After valve actuation, gas flows through the front block via the inlet and channel, as depicted in Fig. 1(b), and into the reservoir. The axially extended nozzle is formed by precise alignment of the two blocks. The de Laval nozzle cross sectional contour is designed using the design code CONTUR [34] and a Fortran90 realization [35]. The code is initialized with an inviscid contour with continuous curvature, which is then corrected for boundary layer growth to achieve uniform flow at the nozzle exit. The design is based on flow Mach number $M = 5$ and throat width 200 μm.

The highly eccentric jet of Fig. 1 does not easily lend itself to the usual interferometric measurement methods applied to rotationally asymmetric jets. One such method is tomography, achieved through multiple interferometric measurements by rotating the jet through a sequence of angles [27,36,37]. For a meter-scale jet, aside from the challenge of rotating it in a typically space-constricted vacuum chamber, the need for a very large field of view would require axially concatenated measurements, making tomography a difficult task. As an alternative, we use a longitudinal interferometric probe beam that propagates along the 30-cm-long nozzle to pick up a longitudinally integrated phase shift profile from the gas sheet. The longitudinal distribution of phase shift contributions is proportional to the longitudinal gas density profile. The latter is measured from plasma fluorescence induced by optical field ionization (OFI) driven by a Bessel beam above the jet nozzle orifice [10,20]. Comparing the fluorescence from the gas jet against Bessel beam-induced OFI of static backfills of known pressure yields the axial gas density profile along the Bessel focus [10,20], from which the transverse profile can be extracted for each longitudinal position.

The measurement results are shown in Fig. 2. The gas jet was situated in a vacuum chamber with access windows for optical probing and imaging. The interferometric probe beam was a λ=532 nm CW diode laser, expanded and collimated to 1 cm diameter. The probe beam propagated parallel to the orifice along the length of the gas sheet, accumulating a phase shift over 30 cm. The probe beam emerging at the end of the jet was imaged through a wavefront shearing interferometer onto a CMOS camera. The solenoid valves were connected by individual solid-state relays to 48VDC power supply; the relays and the camera were triggered by a digital delay generator. The camera image was gated by a 100 μs electronic shutter, much shorter than the ~1 ms evolution timescale of the gas flow. The full gas flow evolution was obtained by collecting a sequence of delayed interferograms by stepwise increase of the camera trigger delay with respect to the valve solenoid triggers. At each delay, 5-10 interferograms were taken (with the jet being fired 5-10 times) and their extracted phase profiles [38] were averaged to reduce the effect of camera noise and increase the signal-to-noise ratio.

As the vacuum chamber pump out times were excessively long for hydrogen, for the interferometric measurements we elected to use nitrogen, with ~5 × faster pump out. Using nitrogen as a stand-in for hydrogen is justified by our fluid flow simulations [39], which show a small difference in reservoir and nozzle flow characteristics for the same inlet pressure. The major difference is in the boundary layer growth, which results in only slightly different Mach number and flow displacement from the nozzle walls for the same inlet pressure. Our simulations show that the maximum density difference between hydrogen and nitrogen gas jets is < ~15% for an inlet pressure 3.4 bar. Furthermore, the time evolution of the gas flow should be very similar for $H_2$ and $N_2$: the rise time is the ratio of reservoir volume to volumetric flow rate out of the nozzle; the latter is similar for the same inlet pressure because the turbulent boundary layer in the 2 mm diameter gas flow channel (Fig. 1(b)) for both $H_2$ and $N_2$ is comparatively negligible.

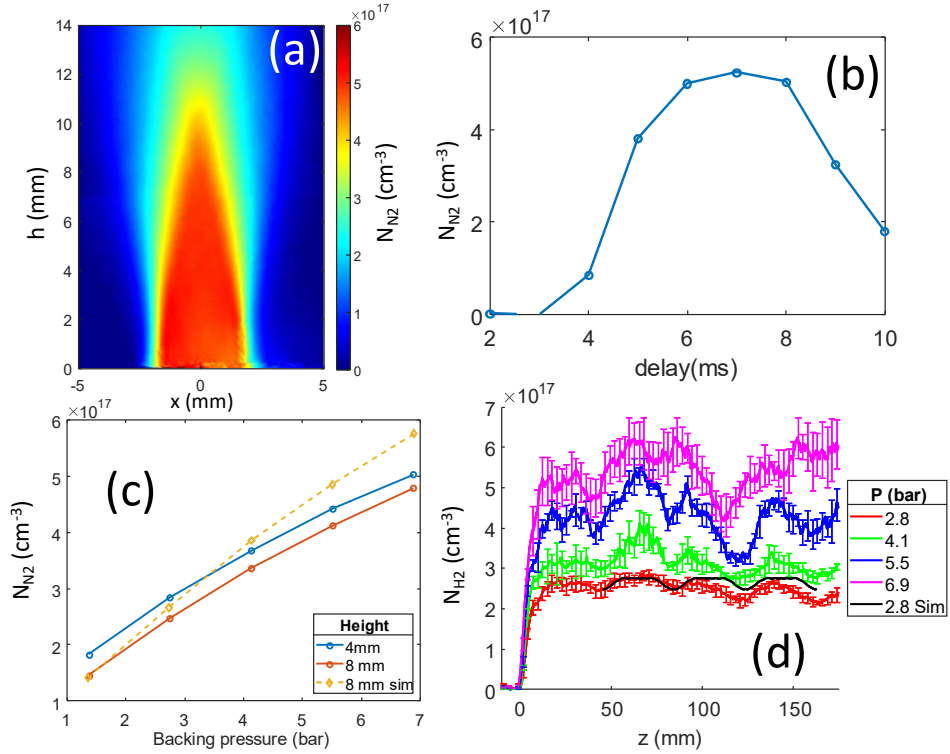

**Figure 2.** Transverse and longitudinal density profile measurements of the 30-cm supersonic gas jet. **(a)** Longitudinal average $N_2$ transverse density profile at 6.9 bar backing pressure. **(b)** Temporal evolution of $N_2$ density $h = 5$ mm above the nozzle orifice at 6.9 bar backing pressure. **(c)** Longitudinal average $N_2$ density versus backing pressure at $h = 4$ mm (blue) and 8 mm (red) above the nozzle orifice. Fluid simulation result ("sim") at $h = 8$ mm is shown for comparison. **(d)** Axially resolved $H_2$ gas density profile at $h = 8$ mm above the nozzle orifice. Fluorescence from approximately one-half the jet length was imaged to construct these profiles. The other half is approximately symmetric owing to the equidistant valve spacing and the reservoir and nozzle symmetry. The result of a fluid flow simulation (see text) is shown for comparison.

The integrated probe phase shift enables determination of the axially averaged transverse gas density profile. An example of an average density profile is shown in Fig. 2(a), for a valve backing pressure of 6.9 bar $N_2$. The gas density is sufficiently low that the probe phase fronts undergo negligible refractive curvature over 30 cm. The axially averaged transverse gas density profile is then calculated as $N(x, h) = N_{STP}\Delta\phi(x, h)/(k_{pr}\Delta n_{STP}L)$, where $\Delta\phi(x, h)$ is the integrated phase shift over $L = 30$ cm, $x$ and $h$ are horizontal and vertical transverse positions with respect to the center of the jet orifice, $k_{pr}$ is the probe wavevector, $(1 + \Delta n_{STP})$ is the refractive index of nitrogen at STP, and $N_{STP}$ is the density of $N_2$ at STP. Owing to the supersonic flow, the central gas density ($x{\sim}0$) is uniform up to $h{\sim}9$ mm above the nozzle orifice; the intent of the supersonic nozzle is to vertically extend the gas flow to enable Bessel beam-induced OFI at heights well above the orifice to avoid clipping of the Bessel beam by front face of the jet. For $h{\sim}5$ mm, the plot in Fig. 2(b) shows the gas density rise time to be ${\sim}2$ ms. The gas density plateaus for ${\sim}2.5$ ms before dropping as the reservoir is depleted. In general, a fast rise time to the desired density is desirable for applications such as LWFA, where excessive gas buildup in the vacuum chamber can refract or distort a high intensity pulse before it reaches the gas jet. In Fig. 2(c), the gas density at two heights $h$ above the nozzle orifice is shown to increase sub-linearly with backing pressure. We attribute this to partial opening of the valves at nearly maximum backing pressure.

The longitudinal gas density profile was measured by imaging the plasma fluorescence. Here, hydrogen was used as the working gas because the fluorescence image is bright enough for a single-shot image. OFI is induced by a 50 fs zeroth order Bessel beam whose central axis was placed in a range 3 to 12 mm above the orifice. The fluorescence intensity from hydrogen recombination is imaged by an sCMOS camera outside the vacuum chamber [10,20] with a H-α line filter (λ=656.3 nm). Images were for one-half of the jet length owing to limitation of the field of view by the chamber viewport aperture. The other half is approximately symmetric owing to the equidistant valve spacing and the reservoir and nozzle symmetry. The longitudinally-resolved $H_2$ density profiles at 8 mm above the nozzle orifice are plotted in Fig. 2(d) for various backing pressures. These profiles are consistent with the longitudinally averaged densities from interferometry plotted in Fig. 2(c). The axial variation in density in each of the curves is from slight nozzle throat width variations from machining and block assembly.

To assist the jet and nozzle design and to gain insight into the flow, we conducted steady state 3D fluid simulations using Ansys Fluent with the SST k-ω model [39]. The flow domain consists of the 19-mm long internal volume shown in Fig. 1(b) and a 19 mm × 55mm × 100 mm (L×W×H) rectangular block above the orifice as the vacuum chamber. The boundary condition at the two ends is continuity of all flow quantities (all gradients equal zero at the ends). The pressure at the inlet (shown in Fig. 1(b)) used for the simulation is determined by matching the inlet flow rate with the Parker C21 flow curve for the experimental valve backing pressure. The black curve in Fig. 2(d) is the simulated density at $h = 8mm$. The periodic curve is a concatenation of identical simulation results (each period is comprised of joined mirror image simulations of Fig. 1(b)), where the density maximum is directly above the gas flow channel and the minima are farthest from the channel. The simulated longitudinal average gas density as a function of valve backing pressure is plotted in Fig. 2(c), with up to ~20% deviation from the measurement. The longitudinal density profile is overlayed in black in Fig. 2(d) for the case of 2.8 bar inlet pressure.

### III. Modular supersonic gas jet

For gas jets longer than a few tens of centimeters, assembling them from modules is highly desirable. First, maintaining fabrication tolerances for long single-module jets is challenging. Second, well-characterized standard modules fabricated to high tolerance can be assembled to construct gas jets of any length. And third, modular jets can enable axial control of gas density and composition. For example, in LWFA a density upramp (an increase in plasma density in the guided beam direction) helps to phase match the accelerated electron bunch with the plasma wake [23]. In electron beam-driven wakefield accelerators, a precise density upramp improves beam focusing while maintaining beam emittance [40–43]. Custom shaped gas jets are also used for enhancing betatron generation [3], long wavelength infrared generation [44,45] and "photon acceleration" [46]. Recent work shows that localized dopant gas for ionization injection in LWFA will improve injected charge and reduce energy spread [47].

To achieve these goals, we developed the first modular design of a meter-scale supersonic slit nozzle and tested its operation. See Fig. 3(a), which is a drawing of two linked modules. Each 11-cm-long module is mainly comprised of two opposing blocks that join together to form the reservoir and nozzle, as in the single module jet of Fig. 1. The internal volume is similar to the de Laval nozzle profile of Fig. 1(b). Each module is divided into 3 independent nozzle sections which can be separated by a custom-designed dividers aimed at various applications, with each nozzle section fed by its own solenoid valve. Each module also has ports for auxiliary valves for dopant gas injection (not shown). The valves share the same gas supply from a common manifold (comprised of parts additional to the nozzle blocks) that comes together upon multi-module assembly. Each added module, which extends the nozzle and manifold, is connected end-on and gasket-sealed to extend the jet to arbitrary lengths.

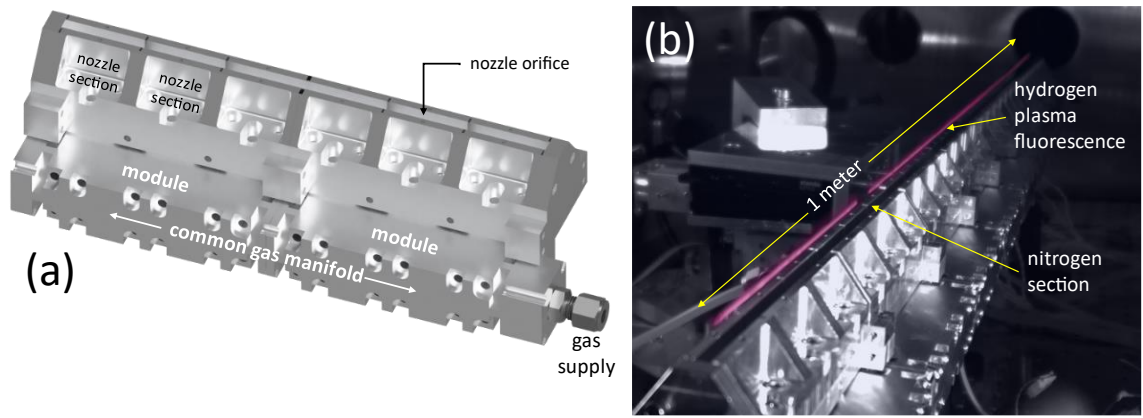

**Figure 3.** (a) 3D model of a two-module supersonic gas jet, including nozzles, manifolds and valves (behind the manifolds). (b) A 1-meter modular supersonic gas jet in use in vacuum. One of the 27 sections is filled with nitrogen and hydrogen elsewhere. An incoming laser pulse is focused by a diffractive logarithmic axicon to form a 1-meter focus above the gas jet. The plasma fluorescence is captured with an H-α line filter where the nitrogen section is shown as a gap.

We have explored control of the axial density profile by adjusting either valve timing (transient control) or valve flow rates (steady state control). In the former case, all valves share the same power supply, while the valve trigger is delayed to adjust the flow rate on a sub-millisecond timescale. In the latter case, individual valve flow rates are controlled by voltage peak or pulse width modulation. The two methods can be combined to achieve more flexible control or reduce the gas load. In this paper we demonstrate the first method.

We characterized the axial $H_2$ density profile of a two-module gas jet (Fig. 4(a)) and the density rise to steady state for a single nozzle section (Fig. 4(b)), using the fluorescence method [10,20] used in Fig. 2. No dividers were used between the nozzle sections within each module. Plots of ramped longitudinal density profiles, controlled by valve trigger timing, at various distances above the nozzle orifice are shown in Fig. 4(a), where the variations about the local mean are from axial variations in the nozzle throat width. The sharp dip at $z = 110$ mm is due to the obstruction of a 125-μm thick shim placed between the two 11-cm-long modules to delineate the joint location. The required trigger timings in Fig. 4(a) are informed by the measured density evolution with respect to the valve trigger of a single nozzle section, as plotted in Fig. 4(b). Earlier than ~3 ms, there is negligible fluorescence signal, and the signal reaches steady state at ~4.5 ms. The transition to steady state takes ~1 ms with a nearly linear temporal ramp. For these conditions, control of axial density profiles is thus possible using temporally staggered valve triggers with inter-trigger intervals as short as ~100 μs. The development of a valve control algorithm is in progress.

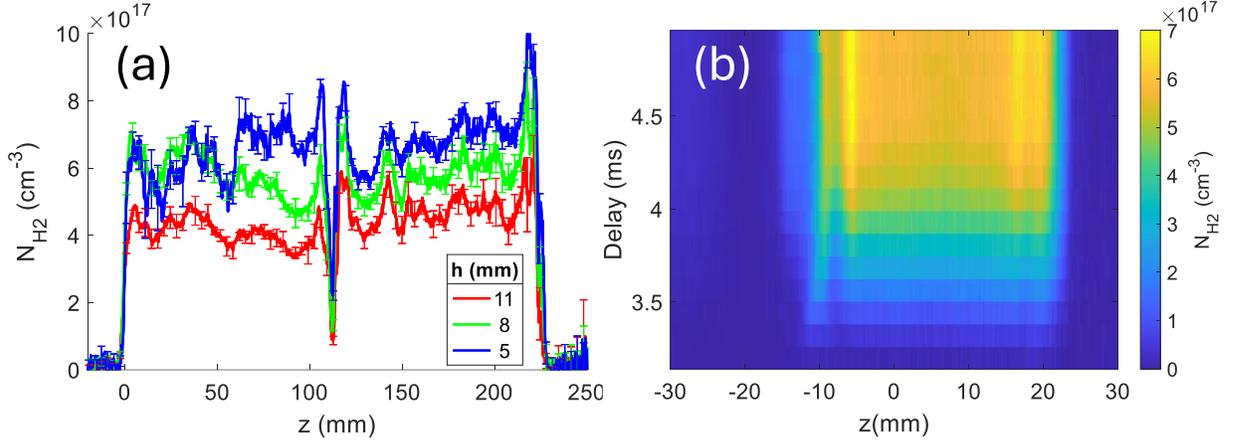

**Figure 4.** **(a)** Longitudinal $H_2$ density profile of a two-module gas jet after steady state is reached. The valve backing pressure is 5.5 bar and each valve is driven by a 48VDC, 4-ms square wave pulse. **(b)** Temporal evolution of a single nozzle section taken at $h = 8$ mm above the nozzle orifice. The valve operating conditions are the same as Fig. 4(a).

Finally, as a proof of engineering feasibility, we assembled a 1.0-meter-long prototype gas jet from 9 modules of the type illustrated in Fig. 3(a) and successfully tested it at the ALEPH laser facility [48] at Colorado State University. Figure 3(b) shows the $H_2$ gas jet in operation in the vacuum chamber. A 100 mJ, 50 fs Bessel beam [49] formed a 1-meter-long OFI hydrogen plasma, 15 mm over the jet orifice. An auxiliary injector valve at one nozzle section was backed by nitrogen, with all other valves backed by hydrogen. The plasma fluorescence image was taken through H-α line filter so that the nitrogen section appears as a gap.

## IV.    Conclusions

We have demonstrated two types of meter-scale supersonic gas jets aimed at LWFA applications. The 30-cm-long single-module version has a reasonably uniform axial density distribution up to ~10 mm above the nozzle orifice, making it very suitable for use in high intensity laser facilities. To scale our design for generation of even longer jets while maintaining fabrication tolerances, and enabling control of axial density profiles, we have developed a modular supersonic jet design. Using two linked modules (6 nozzle sections) of total length 22 cm, we have demonstrated valve timing control of axial density ramps. And using 9 linked modules (27 nozzle sections) of total length 1.0 m, we have demonstrated generation of 1-m-long fully ionized hydrogen plasmas. To our knowledge, this is the longest gas jet plasma yet generated. The modular gas jet is a promising tool for laser wakefield generation and electron-beam driven wakefield acceleration.

**ACKNOWLEDGMENTS**

The authors thank Mayank Gupta, Nishchal Tripathi, Frederica Liu for lab assistance and Dr. Sebastian Lorenz, Dr. Gabriele Grittani for technical discussions. This work was supported by the

## AUTHOR DECLARATIONS

### Conflict of Interest

The authors have no conflicts to disclose.

### Author Contributions

All authors contributed equally to all aspects of this paper.

## DATA AVAILABILITY

The data that support the findings of this study are available within the article and also from the corresponding author upon reasonable request.

49. N. Tripathi, J. Shrock, B. Miao, E. Rockafellow, and H. Milchberg, "Plasma waveguide generation with diffractive logarithmic axicon," Bulletin of the American Physical Society (2023).12